
\input phyzzx
\date{February 1993}
\rightline{University of Tokyo preprint UT 630}
\titlepage
\vskip 1cm
\title{Topological Strings, Flat Coordinates and Gravitational
Descendants}
\author {Tohru Eguchi}
\address{Department of Physics, Faculty of Science, University of
Tokyo, Tokyo 113, Japan}
\author{Hiroaki Kanno \footnote{\dagger}
{Fellow of the Nishina Memorial Foundation}}
\address{DAMTP, University of Cambridge,
Cambridge CB3 9EW, England}
\author{Yasuhiko Yamada}
\address{National Laboratory for High Energy Physics (KEK),
Tsukuba, Ibaraki 305, Japan}
\andauthor{Sung-Kil Yang}
\address{Institute of Physics, University of Tsukuba,
Ibaraki 305, Japan}
\abstract{
We discuss physical spectra and correlation functions
of topological minimal models coupled to topological gravity.
We first study the BRST formalism of these theories and show that
their BRST operator $Q=Q_s+Q_v$ can be brought to $Q_s$ by a certain
homotopy operator $U$, $UQU^{-1}=Q_s$ ($Q_s$ and $Q_v$ are the $N=2$
and diffeomorphism BRST operators, respectively).
The reparametrization
(anti)-ghost $b$ mixes with the supercharge operator $G$ under this
transformation. Existence of this transformation enables us to use
matter fields to represent cohomology classes of the operator $Q$.
We explicitly construct gravitational descendants and show that they
generate the higher-order KdV flows. We also evaluate genus-zero
correlation functions and rederive basic recursion relations of
two-dimensional
topological gravity.}
\endpage
\overfullrule=0pt
\def\cmp#1{{\it Comm. Math. Phys.} {\bf #1}}
\def\pl#1{{\it Phys. Lett.} {\bf B#1}}
\def\prl#1{{\it Phys. Rev. Lett.} {\bf #1}}
\def\np#1{{\it Nucl. Phys.} {\bf B#1}}
\def\ijmp#1{{\it Int. J. Mod. Phys.} {\bf A#1}}
\def\mpl#1{{\it Mod. Phys. Lett.} {\bf A#1}}

\REF\VV{E. Verlinde and H. Verlinde, \np{348} (1991) 457.}
\REF\LI{K. Li, \np{354} (1991) 711.}
\REF\L{A. Losev, "Descendants Constructed from Matter Field in
Topological
Landau-Ginzburg Theories Coupled to Topological Gravity",
ITEP preprint
Nov. 1992.}
\REF\BLNW{M. Bershadsky, W. Lerche, D. Nemeshansky and N. Warner,
"Extended $N=2$ Superconformal Structure of Gravity Coupled to
Matter",
preprint CERN-TH.6694/92, Oct. (1992).}
\REF\D{R. Dijkgraaf, a talk at Newton Institute for Mathematical
Sciences, Dec. 1992.}
\REF\DN{J. Distler and P. Nelson, \prl{66} (1991) 1955.}
\REF\EKY{T. Eguchi, H. Kanno and S.-K. Yang, \pl{298} (1993) 73.}
\REF\IK{H. Ishikawa and M. Kato, "Equivalence of BRST Cohomologies
for 2-D Black Hole and $c=1$ Liouville Theory",
preprint UT-Komaba/92-11, Nov. 1992.}
\REF\Va{C. Vafa, \mpl{6} (1991) 337.}
\REF\DVV{R. Dijkgraaf, E. Verlinde and H. Verlinde, \np{352}
(1991) 59.}
\REF\SF{K. Saito, Publ. RIMS, Kyoto University {\bf 19} (1983) 1231.}
\REF\BV{B. Bloch and A. Varchenko, \ijmp{7} (1992) 1467.}
\REF\SC{K. Saito, "On the Periods of Primitive Integrals", Harvard
Lecture Notes, 1980.}
\REF\GD{I.M. Gelfand and L. Dikii,
Russian Math. Surveys {\bf 30}(5) (1975) 77.}
\REF\BDSS{T. Banks, M. Douglas, N. Seiberg and S. Shenker, \pl{238}
(1990) 279.}
\REF\GM{D. Gross and A. Migdal, \np{340} (1990) 333.}
\REF\DW{R. Dijkgraaf and E. Witten, \np{342} (1990) 486.}
\REF\W{E. Witten, \np{340} (1999) 281.}
\REF\DNI{J. Distler and P. Nelson, \cmp{138} (1991) 273, \np{366}
(1991) 255.}

In this article we will discuss physical spectra and correlation
functions of topological minimal models coupled to topological
gravity.
We first study the BRST formalism of these theories
[\VV,\LI] and show that
their total BRST operator $Q=Q_s +Q_v$ ($Q_s$ and $Q_v$ are the
$N=2$ and diffeomorphism BRST operator, respectively) can be brought
to $Q_s$ by a
certain similarity transformation $U$, $U(Q_s +Q_v)U^{-1}=Q_s$.
This shows that $Q_s$ and $Q=Q_s+Q_v$ have the same spectra and
the chiral primary fields of the matter sector remain physical
observables
after coupling to gravity. In the "matter picture" defined by a
rotation by the operator $U$, the reparametrization (anti-)ghost
field $b$ is replaced by $UbU^{-1}=b+G$ where $G$ is the supercharge
operator. Thus the equivariance condition of the closed string theory
$b_0^- |{\rm phys}\rangle=0$ now reads
$(b_0^- +G_0^-)|{\rm phys}\rangle=0$.
Therefore some of the BRST exact states in the matter theory
$Q_s|\Lambda \rangle$ no longer decouple when
$(b_0^- +G_0^-)|\Lambda\rangle \not=
0$ and will describe gravitational descendant states.

Inspired by the recent work of
Losev [\L] (for related works, see [\BLNW], [\D]) we explicitly
construct gravitational descendants using the
Landau-Ginzburg description
and show that they precisely generate the higher-order flows
of KdV hierarchy.
We also explicitly evaluate genus zero correlation functions and
rederive the basic recursion relations of two-dimensional
topological gravity.

Let us first recall the BRST machinery for the topological gravity
theory developed in refs.[\VV,\LI].
 One first introduces the topological $N=2$ algebra generated by
the stress tensor $T_i$, supercharge operators $G_i^*, \ G_i$ and
the $U(1)$ current $J_i$ for each matter, Liouville
and ghost sector ($i=m,L,gh$, respectively).
The integral of $G_i^*$ is regarded as the
BRST operator and one has
$$
T_i(w)=[ \oint dz G_i^*(z), \ G_i(w)], \ \ \ i=m, \ L, \ gh.
\eqno\eq
$$
The central charge of each sector vanishes separately.
Then the total stress tensor $T$, BRST operator $Q_s$ and the
supercharge operator are defined by
$$
\eqalign{ T &=T_m+T_L+T_{gh} \ ,  \cr
Q_s &=\oint G_m^* +G_L^* +G_{gh}^*  \ ,   \cr
G   &=G_m +G_L +G_{gh} \ .  \cr}
\eqno\eq
$$
$T, \ G^*, \ G$ for the ghost fields are given by
$$
\eqalign{T_{gh} &= -2b \cdot \partial c-\partial b \cdot c
                 -2\beta \cdot \partial \gamma
                 -\partial \beta \cdot \gamma  \cr
G_{gh}^* &=b \gamma  \cr
G_{gh}   &=-c \cdot \partial \beta -2\partial c \cdot \beta~. \cr}
\eqno\eq
$$
In the formulation of ref.[\VV]
the Liouville sector is described in terms of free fields
$\phi, \ \pi, \ \psi, \ \chi,$
$$
\eqalign{T_L &=\partial \pi \partial \phi -\partial^2 \pi
       +\partial \chi \partial \psi    \cr
G_L^* &=\partial \pi \cdot \psi+\partial \psi  \cr
G_L   &=-\partial \chi \partial \phi +\partial^2 \chi~.  \cr}
\eqno\eq
$$
The BRST operator $Q_v$ for the diffeomorphism invariance is
defined by
$$
Q_v =\oint c(z)\tilde T(z)dz -\oint \gamma (z)\tilde G(z)dz.
\eqno\eq
$$
where
$$
\eqalign{
&\tilde T=T_m+T_L+{1 \over 2}T_{gh}, \cr
&\tilde G=G_m+G_L+{1 \over 2}G_{gh}. \cr}
\eqno\eq
$$
It is easy to show that
$$
Q_s^2=Q_v^2= \{ Q_s, \ Q_v \} =0
\eqno\eq
$$
and hence
$$
Q=Q_s+Q_v
\eqno\eq
$$
is nilpotent $Q^2=0$. The above construction
of the BRST operators follows naturally
from a constrained $N=2$ supergravity theory on the Riemann surface
[\DN].

Let us now introduce the following operator
$$
U=\exp [-\oint c(z)\tilde G(z)dz ]~.
\eqno\eq
$$
After some simple algebra one finds
$$
\eqalign{
[\oint c(z)\tilde G(z) dz, \ Q_s] &= Q_v  \cr
[\oint c(z)\tilde G(z) dz, \ Q_v] &= 0 ~. \cr}
\eqno\eq
$$
Hence
$$
U(Q_s +Q_v)U^{-1} =Q_s  \  .
\eqno\eq
$$
We also note
$$
\eqalign{
UbU^{-1} &= b+G  \cr
UGU^{-1} &= G, \ \ \ \ UcU^{-1}=c  \cr
U\beta U^{-1} &= \beta , \ \ \ \
U\gamma U^{-1}=\gamma -c\partial c \cr
U\gamma_0 U^{-1} &= \gamma_0  \cr}
\eqno\eq
$$
where $\gamma_0$ is the basic BRST invariant of the Liouville-ghost
sector
$$
\gamma_0
= {1 \over 2} \big(\{Q_s, \partial c +c\partial \phi \}-c.c. \big)~.
\eqno\eq
$$
The stress-tensor itself is invariant under transformation
$$
UTU^{-1} =T ~ .
\eqno\eq
$$
Let us call the states obtained after the rotation by $U$
as those of the "matter picture"
$$
|{\rm states}\rangle_m =U|{\rm states}\rangle,
\eqno\eq
$$
where $|{\rm states}\rangle$ are in the original picture.
The physical state condition in the matter picture is given by
$$
Q_s |{\rm state}\rangle_m =0,
\eqno\eq
$$
and
$$
(b_0 +G_0)^- |{\rm state}\rangle_m =0~.
\eqno\eq
$$
(17) corresponds to the standard semi-relative cohomology
condition
$(b_0-{\overline b}_0)|{\rm state}\rangle
\equiv b_0^- |{\rm state}\rangle =0$. (16) shows that the $N=2$
chiral primary fields in the matter sector
$Q_s |{\rm chiral}\rangle_m =0$
remain physical after coupling to gravity
$Q \big( U^{-1} |{\rm chiral}\rangle_m \big) =0$, provided
$(b_0 +G_0)^- |{\rm chiral}\rangle_m=0$. $U^{-1}$ generates
certain shifts in the
operators which appear in the chiral field. Furthermore some of
the BRST-exact states $|{\rm state}\rangle_m =Q_s |\Lambda\rangle_m$
become
non-trivial after coupling to gravity if
$(b_0 +G_0)^- |\Lambda\rangle_m \not= 0$.
These are the gravitational descendant states.
Transformation of the BRST operator and shifts in the fields are
reminiscent
of a similar construction which appeared in the study of the
two-dimensional black hole [\EKY], [\IK].

In order to describe the matter sector of the theory we use the
formalism of the topological Landau-Ginzburg (LG) theory
[\Va]
which seems particularly suited to the discussion of gravitational
descendant fields. Let us consider the A-type minimal model at
level $k$. The (perturbed) superpotential is given by
[\DVV]
$$
W(x,t_0, \cdots ,t_k)={1 \over k+2} x^{k+2}
+ \sum_{i=0}^k \; g_i(t)x^i .
\eqno\eq
$$
Perturbation parameters $t_0, \cdots ,t_k$ are chosen to be the
flat coordinates
[\SF,\BV]
of the space of deformations of the superpotential.
They are coupled to the primary fields as
$$
\eqalign{
&\phi_i(x,t)= {\partial \over \partial t_i}W(x,t) \cr
&=\det  \pmatrix{x        &-1         &0      &\ldots   &0      \cr
           t_{k}     &x         &-1      &\ldots   &0      \cr
           t_{k-1}   &t_{k}     &x      &\ldots   &0      \cr
           \vdots    &\vdots    &\vdots &\ddots   &\vdots \cr
           t_{k-i+2} &t_{k-i+3} &\ldots &t_{k}    &x      \cr} \cr
&=x^i + \; \hbox{lower order terms} \ .\cr}
\eqno\eq
$$
The explicit form of the superpotential is obtained by integrating
$\phi_{k+1}=\partial_xW(x)$.
$g_i(t)~ (i=1,2,\cdots,k)$ does not
depend on $t_0$ while $g_0(t)$ has a form
$g_0=t_0+\tilde g_0(t_1,t_2,\cdots,t_k)$.
The metric on the deformation space given by
$$
\eta_{ij} = \langle \phi_i \phi_j \rangle
=\oint dx {\phi_i(x,t) \phi_j(x,t) \over
\partial_x W(x,t)} = \delta_{i+j,k}
\eqno\eq
$$
is independent of the parameters $t_\ell$.
The fusion coefficient $c_{ij\ell}$ is defined as usual by
$$
c_{ij\ell } = \langle \phi_i \phi_j \phi_{\ell} \rangle \ .
\eqno\eq
$$

Let us next introduce the $(k+2)$-th root of the
superpotential [\DVV]
$$
W={1 \over k+2} L^{k+2} .
\eqno\eq
$$
Then the primary field (19) is also expressed as
$$
\phi_i =[L^i \partial_x L]_+ \ ,  \ \ \ i=0,1,\cdots ,k,
\eqno\eq
$$
where $[L^i \partial_x L]_+$ means taking the non-negative powers
of the Laurent
series $L^i \partial_x L$. One can check that (23) reproduces (20)
$$
\oint dx {[L^i \partial_x L]_+ [L^j \partial_x L]_+
           \over \partial_x W}
=\oint dx L^{i+j-k-1} \partial_x L
=\delta_{i+j,k} \
\eqno\eq
$$
where the residue is taken at $\infty$.
We note an important relation
$$
\eqalign{
{\partial \over \partial t_j} \phi_i
&={\partial \over \partial t_j} [L^i \partial_x L]_+
 =\partial_x [L^i {\partial \over \partial t_j} L]_+  \cr
&=\partial_x [L^{i-k-1} \phi_j]_+
 =\partial_x [{L^i \partial_x L \over \partial_x W} \phi_j]_+  \cr
&= [{\phi_i \phi_j \over W'}]_+'~.  \cr}
\eqno\eq
$$
Here the prime means the derivative with respect to $x$.
Following Losev [\L] we introduce the notation
$$
\big[ {\phi_i \phi_j \over W'} \big]'_+ \equiv c(\phi_i, \phi_j) \ .
\eqno\eq
$$
The operation $[~/W']_+'$ originally appeared in the work of K. Saito
in his study of singularity theory [\SC].
When the system is coupled to gravity (20) is replaced by
$$
\eta_{ij} = \langle \phi_i \phi_j P \rangle
\eqno\eq
$$
with $P$ being the puncture operator (we consider correlation
functions on the sphere). The genus-zero three-point function
is evaluated easily in the Landau-Ginzburg theory as
$$
\langle \phi_i \phi_j \phi_\ell \rangle
=\oint dx {\phi_i(x,t) \phi_j(x,t) \phi_\ell (x,t)
\over W'(x,t)} \ .
\eqno\eq
$$

Let us now generalize the range of $i$ in (23) and define the
$n$-th gravitational descendant of the primary field $\phi_i$ by
$$
\eqalign{
&\sigma_n (\phi_i)
\equiv N_{n,i} [L^{(k+2)n+i} \partial_x L]_+ \ ,
          \ \ \ i=0,1,\cdots ,k, \ \ n=0,1,\cdots   \cr
&N_{n,i}=
\big((i+1)(i+1+k+2) \cdots (i+1+(n-1)(k+2))\big)^{-1} \ , \ \
(N_{0,i}=1)~.  \cr}
\eqno\eq
$$
It turns out that these are the fields coupled to the higher-order
flows of the KdV hierarchy.
We note an identity
$$
\eqalign{
\sigma_n (\phi_i)
&= N_{n,i} [L^{i+1+(n-1)(k+2)} \partial_x W]_+ \cr
&= N_{n,i} [L^{i+1+(n-1)(k+2)}]_+ W'
  +N_{n,i} [[L^{i+1+(n-1)(k+2)}]_- W']_+ ~ . \cr}
\eqno\eq
$$
The first term in (30) is equal to
$$
W'(x) \int^x \sigma_{n-1} (\phi_i)
\eqno\eq
$$
while the second term has a degree less than $k+1$ in $x$ and
hence is
expanded into a sum of primary fields
$N_{n,i}[[L^{i+1+(n-1)(k+2)}]_- W']_+
= \sum_{\ell =0}^k c_\ell \phi_\ell$.
Coefficients $c_\ell$ are determined by evaluating
$\langle \sigma_n (\phi_i) P \phi_j \rangle
=\sum c_\ell \langle \phi_\ell \phi_j P \rangle
=\sum c_\ell \eta_{\ell, j}$. Thus
$$
\eqalign{
c_{k-j}
&= N_{n,i} \oint dx {[L^{i+n(k+2)} \partial_x L]_+ \phi_j \over W'}
   \cr
&= N_{n,i} \oint dx L^{i+1+(n-1)(k+2)} \phi_j  \cr
&=N_{n,i} \oint dx L^{i+n(k+2)} \partial_j L  \cr
&=N_{n+1,i} {\partial \over \partial t_j}
{\rm res}[L^{i+1+n(k+2)}]~.  \cr}
\eqno\eq
$$
$N_{n+1,i} {\rm res}[L^{i+1+n(k+2)}]$ is the Gelfand-Dikii
potential $R_{n,i}$ of the KdV hierarchy
[\GD].
Hence we obtain a basic formula for the gravitational descendants

$$
\sigma_n (\phi_i)
=W'(x) \int^x \sigma_{n-1} (\phi_i)
+\sum_{\ell =0}^k {\partial \over \partial t_\ell}
R_{n,i} \phi_{k-\ell} \ .
\eqno\eq
$$
(33) obeys Saito's recursion relation [\SC]
$$
\Big[ {\sigma_n (\phi_i) \over W'} \Big]_+' = \sigma_{n-1} (\phi_i)~.
\eqno\eq
$$

We first note that our formula for gravitational descendants (33)
contains
primary-field components and differs from Losev's expression
$$
\sigma^{L}_n (\phi_i) =
W'(x)\int^x dx_n W'(x_n)\int^{x_n} dx_{n-1} W'(x_{n-1}) \cdots
\int^{x_2} dx_1 \phi_i(x_1)~.
\eqno\eq
$$
Being proportional to $W'$ (35) decouples from the
all three-point functions and hence can not be
a scaling field coupled to the higher-order KdV flow.

By integrating over $t_{\ell}$ an expectation value of
(33), $\langle \sigma_n (\phi_i) P\phi_\ell \rangle
={\partial \over \partial t_\ell} R_{n,i}$,  we also find
$$
 \langle \sigma_n (\phi_i) P \rangle= R_{n,i} ~.
\eqno\eq
$$
(36) gives the standard identification
of the Gelfand-Dikii potentials in terms of the two-point
function of scaling operators
[\BDSS][\GM][\DW].

Let us now recall the BRST transformation laws of the topological
LG theory [\Va]
$$
\eqalign{
\delta x &=0, ~~~~~  \delta x^* = \psi^+    \cr
\delta \psi^- &=2{dW \over dx}, ~~~ \delta \psi^+ =0 \cr
\delta \chi_z &= -\partial x, ~~~
\delta \chi_{\overline{z}} = - \overline{\partial} x \cr}
\eqno\eq
$$
(37) leaves the Lagrangian
$$
\eqalign{
{\cal L} &=\partial x \bar \partial x^*
+\chi_z \bar \partial \psi_L
+\chi_{\bar z}  \partial \psi_R   \cr
&+|{\partial W \over \partial x}|^2
+\psi_R \psi_L {\partial^2 W \over \partial x^{* 2}}
+ \chi_z \chi_{\bar z} {\partial^2 W \over \partial x^2} \cr}
\eqno\eq
$$
invariant ($\psi^\pm = \psi_L \pm \psi_R$) up to a total derivative.
Thus the piece
proportional to $W'$ in (33) is BRST-exact
$$
W'(x)\int^x \sigma_{n-1}(\phi_i) (y)dy
={1 \over 2}Q \big( \psi^- \int^x \sigma_{n-1}(\phi_i) (y)dy \big)
\eqno\eq
$$
and would  decouple in a pure matter theory. In the presence of
gravity, however, one has to examine the equivariance condition.
Using the
representation $G= \chi_z \partial x^*$, one finds
$$
(b_0 +G_0)^- \Big( \psi^- \int^x \sigma_{n-1}(\phi_i) \Big)
= -2 \sigma_{n-1}(\phi_i) \not= 0 ~.
\eqno\eq
$$
Thus the BRST-exact term is non-trivial after coupling to gravity
and the descendant states become independent physical observables.
The importance of the equivariance condition has been emphasized by
Dijkgraaf [\D].

We note that in the critical limit with $t_i=0 ~(i=0,\cdots,k)$,
$\sigma_1(P)=W(x)'x$ and thus
$$
\sigma_1(P)+\gamma_0
={1 \over 2} Q_s \big( \psi^{-}x+(\partial c+c \partial \phi
-{\overline \partial}c - c {\overline \partial}\phi) \big).
\eqno\eq
$$
We find
$$(b_0+G_0)^{-} \big( \psi^{-}x+(\partial c+c \partial \phi
-{\overline \partial}c - c {\overline \partial}\phi) \big)
=-2+2=0.
\eqno\eq
$$
Therefore $\sigma_1(P)=-\gamma_0$ in the equivariant cohomology.

We now turn to the correlation functions and will present
some sample calculations. First of all by repeating the derivation of
(25) we find that the formula is valid also for the gravitational
descendants
$$
{\partial \over \partial t_j}\sigma_n(\phi_i)=
\big[{\sigma_n(\phi_i)\phi_j \over W'} \big]_+'~~.
\eqno\eq
$$
By putting $j=0$ in (43) we obtain an alternate form of the
recursion relation (34)
$$
{\partial \over \partial t_0}\sigma_n(\phi_i)=\sigma_{n-1}(\phi_i).
\eqno\eq
$$
(44) directly leads to the puncture equation [\DW] for
correlation functions.
For instance, in the case of a 3-point function we have
$$
\eqalign{
&{\partial \over \partial t_0} \langle \sigma_{n_1}(\phi_i)
\sigma_{n_2}(\phi_j)\sigma_{n_3}(\phi_{\ell})\rangle
=\langle P\sigma_{n_1}(\phi_i)
\sigma_{n_2}(\phi_j)\sigma_{n_3}(\phi_{\ell})\rangle \cr
&={\partial \over \partial t_0} \oint dx {\sigma_{n_1}(\phi_i(x))
\sigma_{n_2}(\phi_j(x))\sigma_{n_3}(\phi_{\ell}(x)) \over W'(x)} \cr
&=\langle \sigma_{n_1-1}(\phi_i)
\sigma_{n_2}(\phi_j)\sigma_{n_3}(\phi_{\ell})\rangle+
\langle \sigma_{n_1}(\phi_i)
\sigma_{n_2-1}(\phi_j)\sigma_{n_3}(\phi_{\ell})\rangle \cr
&~~~+\langle \sigma_{n_1}(\phi_i)
\sigma_{n_2}(\phi_j)\sigma_{n_3-1}(\phi_{\ell})\rangle~~. \cr}
\eqno\eq
$$
By combining the puncture equation with (36) we find a formula
for the 1-point function on the
sphere
$$
\eqalign{
\langle \sigma_n (\phi_i) \rangle
&=R_{n+1, i} \cr
&=N_{n+2,i} {\rm res}(L^{(n+1)(k+2)+i+1}) ~. \cr}
\eqno\eq
$$
This is the generalization of a known result [\DVV]
$$
\langle \phi_i \rangle ={1 \over (i+1)(i+k+3)} {\rm res}(L^{k+i+3})~.
\eqno\eq
$$

By using our representation (33) of the gravitational descendants
and (46)
we also find
$$
\eqalign{
&\langle \sigma_n(\phi_i)\phi_j \phi_m \rangle
=\sum_{\ell=0}^{k} {\partial \over \partial t_\ell} R_{n,i}
\langle \phi_{k-\ell}\phi_j\phi_{m}\rangle \cr
&=\sum_{\ell=0}^{k}{\partial \over \partial t_\ell}\langle
\sigma_{n-1}(\phi_i)\rangle
\langle\phi_{k-\ell}\phi_j\phi_{m}\rangle \cr
&=\sum_{\ell=0}^k \langle \sigma_{n-1}(\phi_i)\phi_{\ell}\rangle
\langle\phi^{\ell}\phi_j\phi_{m}\rangle~. \cr}
\eqno\eq
$$
This is the topological recursion relation of Witten
[\W].

By differentiating a 3-point function with respect to
the parameters of the small phase space
$t_m (m=0,1,\dots,k)$ we obtain a 4-point function with an
insertion of a $2$-form operator
$\phi_m^{(2)}$. Using (25) we have
$$
\eqalign{
&\langle \phi_{\ell} \phi_j \sigma_n(\phi_i)
\int_{\Sigma}\phi_m^{(2)}\rangle ={\partial \over \partial t_m}
\langle \phi_{\ell} \phi_j
\sigma_n(\phi_i)\rangle
={\partial \over \partial t_m} \oint dx {\phi_{\ell} \phi_j
\sigma_n(\phi_i) \over W'} \cr
&=-\oint dx
{\phi_{\ell} \phi_j \sigma_n(\phi_i) \phi'_m \over W'^2} \cr
&+\oint dx {1 \over W'} \{c(\phi_m,\phi_\ell) \phi_j \sigma_n(\phi_i)
+c(\phi_m,\phi_j) \phi_\ell \sigma_n(\phi_i)
+c(\phi_m,\sigma_n(\phi_i)) \phi_\ell \phi_j \}~. \cr}
\eqno\eq
$$
In the extreme right-hand-side of (49) the 2nd terms are interpreted
as the contact terms [\L]
which arise when the position of the $2$-form
operator $\phi_m^{(2)}$ collides with those
of the other
fields. A contact term in fact describes the  change of the scaling
fields $\phi_i=\partial_i W$
(wave-function renormalization) due to the variation of the
superpotential under perturbation.
We have explicitly checked the validity of (49) by evaluating
the contour integrals.

It is known [\DVV] that  a 4-point function has a common value
irrespective
of which 3 of the 4 operators are chosen to be the $0$-form (and the
remaining one to be a $2$-form). Thus (49) must be equal to, for
instance, $ \langle \phi_{\ell} \phi_j \phi_m \int_{\Sigma}
\sigma_n^{(2)}(\phi_i) \rangle$. We can check
that the right-hand-side of (49)
is in fact symmetric in $\phi_{\ell},\phi_j,\phi_m$ and
$\sigma_n(\phi_i)$ [\L]. Thus
$$
\eqalign{
&\langle \phi_{\ell} \phi_j \phi_m \int_{\Sigma}
\sigma_n^{(2)}(\phi_i) \rangle
=-\oint dx
{\phi_{\ell} \phi_j \phi_m \sigma_n(\phi_i)'  \over W'^2} \cr
&+\oint dx
{1 \over W'} \{c(\sigma_n(\phi_i),\phi_\ell) \phi_j \phi_m
+c(\sigma_n(\phi_i),\phi_j) \phi_\ell \phi_m  \cr
&+c(\sigma_n(\phi_i),\phi_m) \phi_\ell \phi_j \}~. \cr}
\eqno\eq
$$
(50) formally has the same structure as (49).

In this paper we confined ourselves to the small-phase space of
two-dimensional gravity. It will be interesting to see if the
Landau-Ginzburg formulation could handle the large-phase space with
non-vanishing coupling constants to the gravitational descendants.

A field theoretic derivation of the structure of the contact term (26)
may be given along the lines of
[\VV,\DNI] and will be discussed elsewhere.

The research of T.E. and S.K.Y. is partly supported by the
Grant-in-Aid for Scientific Research
on Priority Area "Infinite Analysis".

\refout
\end